\documentclass[twocolumn,aps,showpacs]{revtex4}  
\usepackage{graphicx}%
\newcommand{\figwidth}{3.in}
\begin{document}

\title{Hole-pair hopping in arrangements of hole-rich/hole-poor  domains
in a quantum antiferromagnet}
\author{Efstratios  Manousakis}
\affiliation{   \\
 Department of Physics and MARTECH,
Florida State University,
Tallahassee, Florida, 32306, USA and\\
Department of Physics, Solid State Physics Section,
University of Athens, Athens, Greece } 
\date{\today} 

\begin{abstract}
We study the motion of holes in a doped quantum  antiferromagnet 
in the presence of arrangements of hole-rich and hole-poor
domains such as the stripe-phase in high-$T_C$ cuprates.  When
these structures form, it  becomes energetically favorable
for single holes, pairs of holes or small bound-hole clusters to hop from 
one hole-rich domain to another due  to  quantum  fluctuations. 
However, we find that at temperature of approximately $100 K$, 
the probability for 
bound hole-pair  exchange between neighboring hole-rich regions in the
stripe phase, is  one or two orders of magnitude larger 
than single-hole or multi-hole droplet 
exchange. 
 As a result holes in a given hole-rich domain  penetrate
further into the antiferromagnetically aligned domains  when
 they do it in pairs. At temperature $T  \sim 100 K$ and below bound pairs 
of holes hop from one 
hole-rich domain to another with high probability. 
Therefore our main finding is that the presence of the antiferromagnetic 
hole-poor domains act as a filter which selects, from the 
hole-rich domains (where holes form a self-bound liquid), hole 
pairs which can be exchanged throughout the system.
This fluid of bound hole pairs
can undergo a superfluid phase ordering at 
the above mentioned temperature scale.

\end{abstract}
\pacs{74.20.-z,71.10.-w,71.45.Lr}
\maketitle
\section{Introduction}

When an isotropic quantum antiferromagnet on an infinite 
square lattice, as described 
by the $t-J$ model without the
inclusion of the long-range part of the Coulomb interaction, 
is doped with two holes a
bound state is formed\cite{boninsegni}.
This bound state is not a sign of a pairing instability but that of a
phase separation instability\cite{EKL,HM} or possibly stripe 
ordering\cite{stripeorder}.
While in the high-$T_c$ cuprates stripes have been experimentally observed, 
at the theoretical level, there are two different scenarios to 
explain their formation.  In the first scenario, when the
antiferromagnet is doped with more than two holes they form a 
larger droplet and  there is evidence\cite{GFMC,erica}
that even for relatively small values of $J/t$ there is a critical electron 
density $n_{ps}(J/t)$  above which the system is separated into
two phases: one phase in which the holes are bound into a hole-rich region and 
an all electron phase characterized by antiferromagnetic order.
When one adds the long-range part of the Coulomb interaction this phase
is forbidden because of the fact that the energy density of a droplet 
with finite charge density blows up as the droplet size goes to infinity. 
The system can accommodate\cite{kivelson} this tendency for 
phase separation by forming a state which is made out of alternating 
microscopic regions of the two phases but with a zero net charge 
when averaged over a larger region.
Such possible states are shown in Fig.~\ref{stripe} and in Fig.~\ref{hexagon}.
In the second scenario the stripe-ordered state is obtained right away
as the ground state of the $t-J$ model without the inclusion of the 
long-range Coulomb interaction\cite{stripeorder}. The validity of these
calculations has been discussed\cite{boundary} and the issue has 
not been resolved unequivocally yet.

In this paper we consider the electronic system in a state 
characterized by alternating regions
or domains with different charge such as the static or dynamic stripe 
phase (Fig.~\ref{stripe}) or that of Fig.~\ref{hexagon}
or other structures which take care of the tendency for 
phase separation and the effects of long-range Coulomb
repulsion; when such states form, 
it  becomes energetically favorable
for single holes or small bound-hole clusters to hop from 
one hole-rich domain to another due  to  quantum  fluctuations. 
We show that the probability  of bound hole-pair 
hop is one to two orders of magnitude  
larger than that of single-hole 
 or multi-hole droplet hop between neighboring hole-rich regions. 
Therefore, pairs of holes in a given hole-rich domain
prefer to penetrate further into the antiferromagnetically ordered
domains relative to
single or multi-hole clusters and they have 
a much higher chance to tunnel from one hole-rich domain to another. In 
particular, we find that below $T \sim 100 K$ these bound hole-pairs 
tunnel through the antiferromagnetic domains at high rates.
Therefore, the system can be pictured
as consisted of two inter-penetrating subsystems 
a) a subsystem of hole-rich domains in which the holes exist as a self-bound
liquid and 
b) a fluid of bound hole pairs which exists in the antiferromagnetically
ordered hole-poor domains. The latter fluid is 
composed of bound hole-pairs with a large characteristic binding energy 
scale of the order of the phase separation energy or the stripe formation
energy scale. The main point of this paper is that the existence of the
antiferromagnetic domains act as a filter which allows only pairs
to exist over the entire system.

The second part of this paper includes a discussion which 
involves some speculation. The case of strong pairing correlations within each hole-rich strip and
the role of the inter-strip coupling has been examined by
other authors\cite{erica}.
We wish to consider the limit where pairing within
each hole-rich domain is weak at temperature of the order of $100 K$.
We argue that the fluid of bound hole-pairs, which is selected by the 
presence of the intervening antiferromagnetic regions and
exists throughout the system, can undergo a superfluid phase 
ordering at a relatively high critical temperature. In the limit of
weak pairing within the hole-rich domains, this
critical temperature associated with superfluid phase ordering 
depends on the pair 
effective mass inside the antiferromagnetic domain and the distance between
the hole-rich domains.

\section{Single-hole, hole-pair and multi-hole tunneling}
In order to understand single-hole, hole-pair or
multi-hole tunneling from one hole-rich domain (which could be a 
hole-rich strip or droplet) we will use a continuum model of holes
of effective mass $m^*$, interacting through a mean field 
$V_{eff}$ introduced by the environment and the 
stripe-ordered or the dynamic-stripe configuration and via a 
residual hole-hole interaction $V$.
The partition function for the system of $N$ holes can 
be written, using Feynman\cite{Feynman,Feynman2} path integral formulation, 
as follows:
\begin{eqnarray}
Z &=& {1 \over {N!}} \sum_{P} (-1)^{[P]} 
\int_{\vec x_i(\beta)=P \vec x_{i}(0)}  \prod_{i=1}^{N} 
{\cal D} x_i(\tau)
e^{-S_{eff}} \nonumber \\
S_{eff} &=& \int_0^{\beta} d\tau \Biggl [ \sum_{i=1}^{N} 
\Biggl ({ {m^*} \over {2 \hbar^2}} \biggl ({{d \vec x_i(\tau)}\over  
{d \tau}}\biggr )^2 + 
V_{eff}(\vec x_i(\tau)) \Biggr ) \nonumber \\
 &+& \sum_{i<j} V(\vec x_i(\tau) - \vec x_j(\tau)) \Biggr ],
\label{path-integral}
\end{eqnarray}
where the sum is over all $N!$ hole-permutations $P$ and $[P]$ is the order
of permutation. The integration is over  paths $\vec x_i(\tau)$ 
in imaginary time $\tau$.   The final configuration $\vec x_i(\beta)$ at 
$\tau = \beta$ is any permutation 
$ P \vec x_i(0)$ of the initial hole-positions. 

\begin{figure}[htp]
\includegraphics[width=\figwidth]{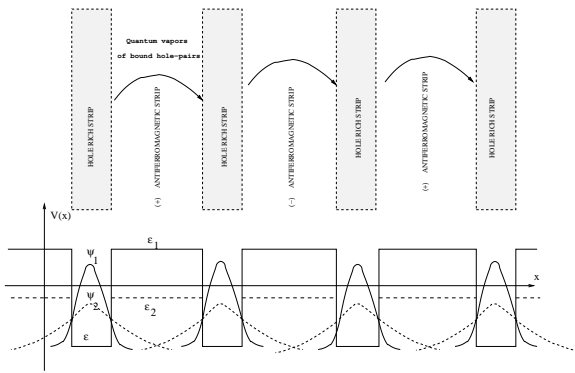}
\caption{(a) A striped ordered state, i.e., alternating strips of the
hole-rich and the hole-poor phases. Two successive hole-poor domains are
separated by a pi-phase shift. In general we consider
the case of dynamic stripes where while these alternating domains
exist there is no stripe ordering. (b) The effective potentials experienced
by a single hole (solid line) and bound hole-pair when attempting to 
tunnel from one hole-rich region to another. The corresponding wave functions
of the single-hole and bound hole-pair are also shown schematically as $\psi_1$ and $\psi_2$. }
\label{stripe}
\end{figure}

\begin{figure}[htp]
\includegraphics[width=\figwidth]{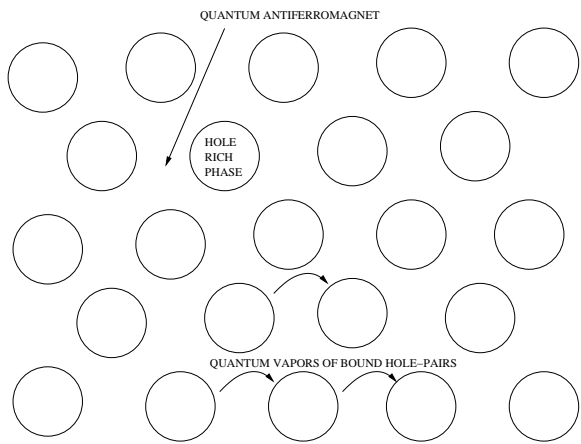}
\caption{A hexagonal superlattice formed from droplets of the
hole-rich phase inside an antiferromagnetic background.}
\label{hexagon}
\end{figure}

Let us consider a state of alternating hole-rich and antiferromagnetically
aligned hole-poor domains. When these alternating regions order they form
the stripe-ordered state shown in Fig.~\ref{stripe}(a).
In addition we will keep in mind other inhomogeneous configurations 
such as the one in Fig.~\ref{hexagon}. In the case of the stripe
state, each hole experiences a potential $V_{eff}$ with a profile shown
in Fig.~\ref{stripe}(b). We wish to understand the 
tunneling of  one or more holes from one hole-rich domain to the 
nearest hole-rich domain. 
First, let us consider the motion of a single-hole and estimate the
factor $exp(-S_{eff})$ for a hole displacement from the boundary of 
a hole-rich domain by a distance $x$ within imaginary time $\tau$ 
inside the antiferromagnetic hole-poor
domain. At the beginning the hole is near the boundary of 
the hole-rich with the hole-poor region and so an estimate 
of the $S_{eff}$ is given as follows
\begin{eqnarray}
S^{(1)}_{eff}(x,\tau) = {{m^*_1} \over {2 \hbar^2}} {{x^2}  \over {\tau}}
+ V_1 \tau.
\label{action1}
\end{eqnarray}
Here $m^*_1$ is the hole effective mass inside the antiferromagnet 
(hole-poor region),
$V_1$ is the difference between the hole energy when the hole is in the 
hole-poor region and the hole energy when the hole is in the hole-rich region.
This is determined by the presence of an effective barrier imposed by the
existence of the two different domains and more significantly by the 
hole-hole interaction.
A single hole sticks to the hole-rich region strongly 
while a pair of holes sticks to the rest of the hole-rich region
relatively weakly. 
This is shown on the basis of the $t-J$ model later
below. This means that the value of the barrier $V_1$ for a single hole to
enter the antiferromagnetic region is significantly larger than the barrier $V_2$ which is experienced by a pair of holes.
The path with the most significant contribution 
to the partition function
is that which corresponds to a characteristic time scale $\tau$ 
(the time for which the hole stays in the hole-poor region) such that
$S^{(1)}(x,\tau)$ for a given $x$ is minimum. The minimum of the
action (\ref{action1}) is obtained for 
\begin{eqnarray}
\tau_0 = x \sqrt { { m^*_1} \over {2 \hbar^2 V_1}}
\end{eqnarray}
and for this value of $\tau$ the action is given by 
\begin{eqnarray}
S^{(1)}_{min} =  \sqrt { { 2 m^*_1 V_1} \over {\hbar^2}} x.
\end{eqnarray}
This gives the well-known result for the quantum mechanical penetration 
probability ($e^{-S^{(1)}}$). Therefore the ``penetration'' depth inside the
hole-poor region (the value of $x$ where $e^{-S^{(1)}} =1/e$) is given by
\begin{equation}
\lambda_1 = {{\hbar} \over {\sqrt{2 m^*_1 V_1} }}.
\label{lambda1}
\end{equation}
As we will show below a pair of holes sticks much less strongly 
to the hole-rich 
region; thus, the difference in energy $V_2$ for the pair
in the hole-rich and hole-poor domains is much smaller than $V_1$. The
 penetration depth for  hole-pair or multi-hole tunneling is given by
\begin{equation}
\lambda_N = {{\hbar} \over {\sqrt{2 m^*_N V_N} }}.
\label{lambdan}
\end{equation}
We would like to estimate these penetration depths and probabilities
on the basis of  results obtained by Green's function Monte Carlo 
simulation of the $t-J$ model\cite{boninsegni1,boninsegni,GFMC,sorella98}.

\begin{figure}[htp]
\includegraphics[width=\figwidth]{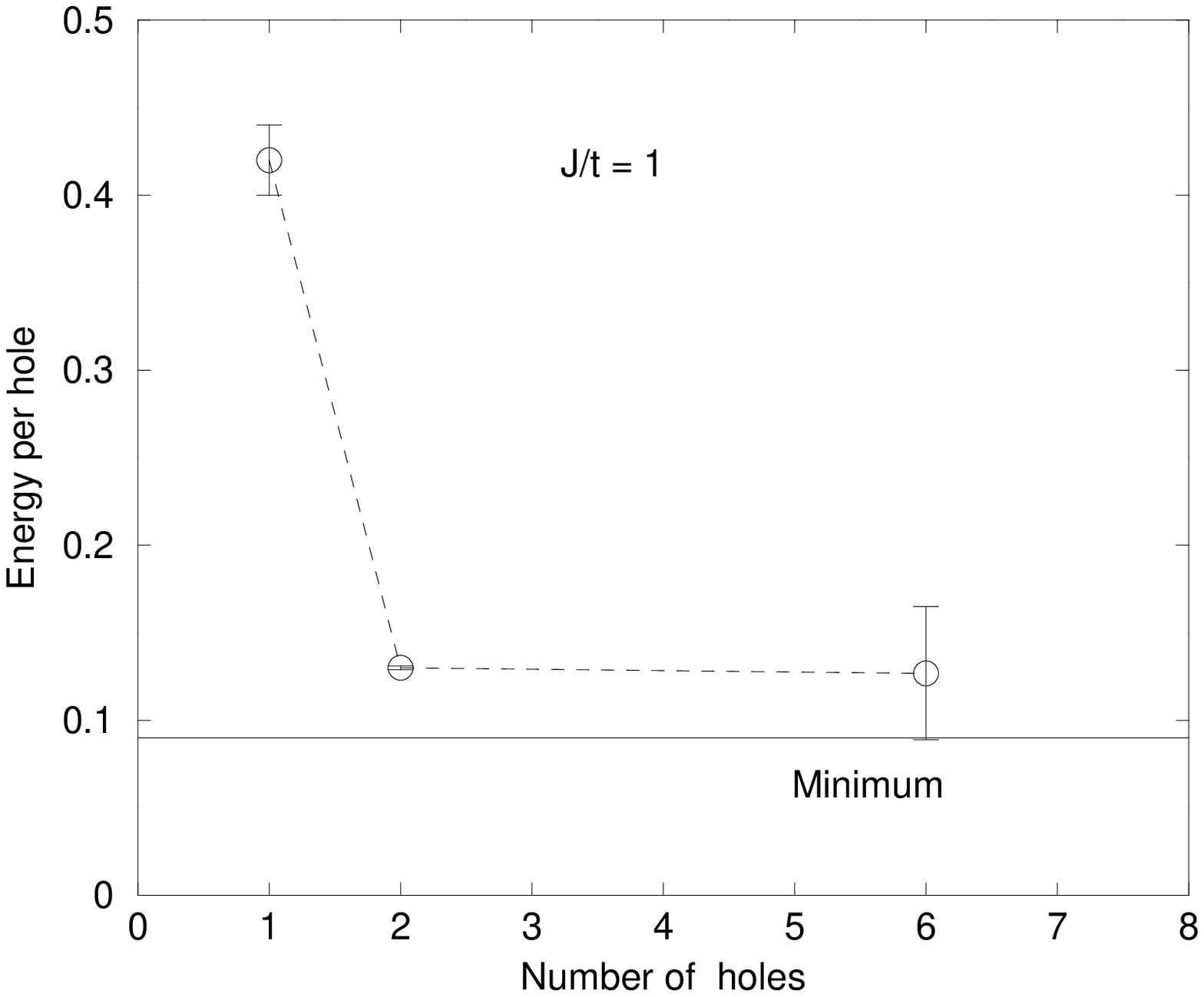}
\caption{The energy per hole of a hole-rich droplet in 
an antiferromagnetic background
as a function of the number of holes at $J/t=1$ 
(dashed line and solid circles).  The solid line is the value of the
minimum value of the energy per hole as a function of the hole density 
for $J/t = 1$}
\label{fig-ehj1}
\end{figure}

\begin{figure}[htp]
\includegraphics[width=\figwidth]{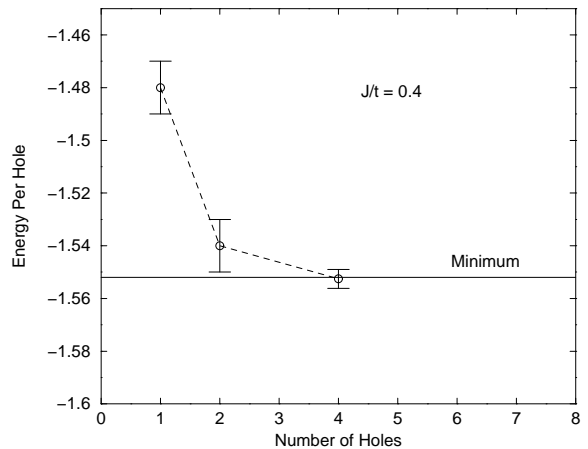}
\caption{The energy per hole of a hole-rich droplet in 
an antiferromagnetic background
as a function of the number of holes at $J/t=0.4$ 
(dashed line and solid circles).  The solid line is the value of the
minimum value of the energy per hole as a function of the hole density 
for $J/t = 0.4$}
\label{fig-ehjp4}
\end{figure}
We consider the $t-J$ model in the parameter range where phase separation
was found\cite{HM,GFMC,sorella98}. Starting from the undoped insulator 
we introduce $N$ holes in a system of $N_s=L\times L$ sites. 
The ground state 
of the system keeping $N$ finite and $L \to \infty$ is expected to 
be a two-dimensional droplet with energy per hole given by 
$e(N)=(E(N)-E(0))/N$. 
The situation is shown in Fig.~\ref{fig-ehj1} for $J/t=1$ and
in Fig.~\ref{fig-ehjp4} for $J/t=0.4$. The circles with the error bars
give the energy per hole $e(N)$ as a function of the number of holes.
The solid line in both figures denotes the minimum of the energy per hole as 
a function of the hole density, namely the energy of the phase separated 
state $e_{ps}$. The energy per hole $e(N)$
is always higher than $e_{ps}$.  The quantity $V_{N}$ discussed previously,
is given by $(e(N)-e_{ps})N$. The important facts to notice is 
that a) the energy per hole for the bound-pair of holes is quite 
close to that of the energy per hole in the phase separated state and b)
introducing two holes inside the quantum antiferromagnet lowers
the energy per hole significantly relative to the single hole case
and further introduction of holes leads to formation of droplets without
as significant gain in binding energy. This implies that it is far easier
to separate a hole-pair than a single-hole from the hole-rich region.
The energy of a hole-rich droplet inside a quantum antiferromagnet
has an additional contribution due to long-range Coulomb repulsion. 
It is clear than when we switch on this interaction, the macroscopically
phase separated state cannot be realized because the energy per hole has 
positive infinite energy. However, a state of microscopic
phase separation is allowed because there is no net charge on a large 
scale and examples of such states are given in Fig.~\ref{stripe}
and Fig.~\ref{hexagon}. The effect of this long-range interaction
does not alter our conclusion and it will be discussed later.

While $m^*_N$ increases almost linearly with the number of holes 
$N$\cite{trugman}, the function
 $V_{N} = (e(N)- e_{ps}) N$ drops sharply at the value $N=2$. 
As a result the combination of $m^*_N V_N$ which occurs in
Eq.~\ref{lambdan} causes a peak in $\lambda_N$ at $N=2$. 
The fact that the largest $\lambda_N$ corresponds to bound hole-pairs
implies that when  stripes form, because of the 
overlap of the wave-function of hole clusters within each hole-rich region with
the  wave-function of a neighboring hole-rich strip, the number of 
bound hole-pairs which will be shared among all strips is much larger
than the number of single holes or the number of any other multi-hole 
clusters shared by all the strips. In addition using  values for
$V_N$ and $m^*_N/m$  obtained from numerical studies of the
$t-J$ model\cite{boninsegni1,boninsegni,GFMC,sorella98} and 
the Hubbard model\cite{trugman} we find that  
$\lambda_2$ is of the order of $10 \AA$ which is of the
inter-strip distance when the stripes order. In the next section we show that 
the contribution of such tunneling configurations 
(in which hole-pairs cross from a hole-rich strip to a neighboring
hole-rich strip through the intervening antiferromagnetic 
hole-poor domains) to the path 
integral (\ref{path-integral}) become very significant at and below a 
temperature of approximately $100 K$.

Tunneling of hole-pairs between neighboring strips contribute
to the path-integral (\ref{path-integral}) by a factor 
$exp(-S_{pair}(d,\beta))$ 
while when a single-hole crosses from one hole-rich strip to a
neighboring one through the antiferromagnetic domain it contributes
to the path integral by an amount $exp(-S_1(d,\beta))$. The ratio
of these contributions is given by $exp(-\Delta S_{12}(d,\beta))$
where $\Delta S_{12}(d,\beta)$ can be estimated as
\begin{eqnarray} 
\Delta S_{12}(d,\beta) =
{{m^*_1-m^*_2} \over {2 \hbar^2}} {{d^2}  \over {\beta}}
+ (V_1-V_2) \beta.
\end{eqnarray}
Taking $T =100 K$  and using the available results for $J/t=0.4$,
namely, $V_1 \sim 0.07 t$,  $V_2 \sim  0.02 t$, and the values of 
$m^*_1/m_0 \sim 6.5, m^*_2/m_0 \sim 22.5 $ 
(as obtained by Trugman\cite{trugman} for $U/t=10$) 
and the value $d\sim 16 \AA$ (a typical distance between 
strips in a stripe-ordered
state)  we find that $S_1-S_{pair} \sim 4$. This implies that 
the contribution of a single hole $e^{-S_1}$ is approximately 1-2 orders 
of magnitude smaller than that of a bound hole-pair at the above mentioned
temperature.

In order to give an accurate determination of such transition rates
for hole-pairs we need a) a more accurate determination
of the pair binding energy difference relative to the energy of the
phase separated state and b) to know the effect of the long-range 
Coulomb repulsion on the binding energy of two holes inside 
an antiferromagnetic domain relative to the energy of these two
holes when they are in the hole-rich domain. The value we used above,
namely $V_2 \sim 0.02 t$ will be significantly reduced in the presence
of the long-range (LR) Coulomb interaction. The turning on of the 
LR Coulomb
interaction will reduce the binding energy of a bound hole-pair
to the rest of the hole-rich region. The stripes result
from the competition between the tendency of the system for 
 phase separation and 
the LR Coulomb interaction. Therefore the hole-rich region contains
just enough holes to keep the balance between these two opposing tendencies.
This implies that the binding energy of a bound pair to the rest of the
hole-rich domain relative to the case where the hole-pair is inside the
antiferromagnetic hole-poor region is very small.  This is in agreement
with the findings of  Arrigoni et al.\cite{arrigoni} who studied the
$t-J$ model and they added the LR Coulomb interaction using a 
combination of density matrix renormalization group method for the
short-range part and a Hartree approximation to take into account the
LR part. When they added the LR Coulomb interaction they find enhanced
superconducting correlations which was associated to the enhancement of
pair tunneling between stripes.  

\section{Superfluidity of the gas of bound hole-pairs}

This section includes a discussion which is speculative. 
If pairing correlations within a given hole-rich stripe domain are strong
 we are in a limit which has
been considered by other authors\cite{erica}. In such case 
a weak inter-strip coupling can give rise 
to a high $T_C$ and superconductivity competes with
charge density wave ordering.
Here we wish to examine the case where pairing correlations 
within a given strip are weak 
at temperature of the order of $100 K$. Strong pairing correlations exist
in ladders but when stripes occur in a two-dimensional system
it is uncertain that such one-dimensional pairing correlations are strong
at this temperature.

If we neglect $V_2$,  
$S_{pair} \sim (m^*_2 d^2/2 \hbar^2) K_B T $ and it is of order 
of unity at $T\sim 100 K$.
At around this temperature the system of such hole-pairs 
acquires long-range phase coherence and non-zero winding number 
due to long chains of
exchanges of hole-pairs between neighboring strips.
The most significant contributions
to the partition function (\ref{path-integral}) at temperature 
of the order of $100 K$ and below  can be written as 
\begin{eqnarray}
Z &=& {1 \over {N_p!}} 
\Bigl ({{m^*_2} \over {2 \pi \hbar^2 \beta}} \Bigr )^{{3 N_p} \over 2}
 \int \prod_{i=1}^{N_p} d\vec z_i  g(\vec z_1,\vec z_2,...,\vec z_{N_p}) 
\nonumber \\
& & \sum_{P} \exp \Biggl (-{ {m^*_2} \over {2 \beta \hbar^2}} \sum_i^{N_p}(\vec z_i - P \vec z_i)^2 \Biggr ).
\label{path-integral2}
\end{eqnarray}
where $N_p=N/2$ is the number of hole pairs in the system with a 
distribution 
governed by the function $g(\vec z_1,\vec z_2,...,\vec z_{N_p})$ giving the 
probability for a given configuration $\{\vec z_i \}$ of the 
bound hole pairs  as in a Bose fluid. There is a sum over all
possible permutations of hole-pairs. Here we
have assumed that the path integration over all trajectories in 
Euclidean-time in the path-integral of Eq. (\ref{path-integral}) 
has been carried out\cite{Feynman2}.  
Let us consider a typical displacement of a hole inside the hole-rich
domains  which is associated with the factor
 $f_1=\exp(-m'_1 |\vec x|^2/2 \hbar^2/\tau - U_1(\tau)) $
where $U_1(\tau)$ is the interaction part of the action and $m'_1$ the
hole effective mass inside the hole-rich domains.
Similarly when a pair of holes moves a similar typical distance
between pairs inside the hole-rich domain we can associate a factor
 $f_2=\exp(-m'_2 |\vec x|^2/2 \hbar^2/\tau - U_2(\tau)) $.
For hole displacements of the order of the 
inter-hole distance in the hole-rich domains, 
both $f_1$ and $f_2$ become of order of unity
for significantly smaller values of the 
imaginary time $\tau$ (higher
temperatures) as compared to the time scale required for hole-pair exchange
between neighboring hole-rich domains.
Namely, in order to provide an optimum hole distribution within 
the strips the holes, in these regions, adjust their positions 
in shorter times scales; the effects discussed previously 
which involve tunneling of 
pairs of holes between two neighboring hole-rich regions are longer-time
effects. Therefore we have assumed that at such lower temperature
where these latter effects become energetically favorable, we can integrate
out these short times scale configurations; 
the holes within each strip have enough time to relax to a distribution   
$g(\vec z_1,...,\vec z_{N_p})$ which is experienced by the bound hole-pairs as
they move from one hole-rich domain to another. This distribution 
$g$ describes the stripe ordering and the charge-stripe correlations.
We have eliminated the single-hole degrees of freedom even within a strip. 
We consider the hole-rich 
domains as made of pairs of holes with the same charge distribution
because what is
important in describing our ideas is the movement of the gas of bound
hole pairs in the space between hole-rich domains. The space occupied by
the hole-rich strips plays only the role of a reservoir from where  
these bound  hole pairs emerge.  The characteristic time-scale 
associated with the motion of these degrees of freedom in 
imaginary time is much shorter than the inverse temperature at which
tunneling of bound hole-pairs between hole-rich domains becomes
likely. 

When a bound hole-pair is exchanged between neighboring stripes, it leads to
a factor $f=\exp(- m^*_2/ (2 \beta \hbar^2) 
(\vec z_i -  \vec z_j)^2) $. This can occur anywhere in the system between
any two neighboring stripes. These pair-hole movements cannot occur in 
an uncorrelated manner because when the pairs move to another hole-rich 
domain they disturb the balance of charge. Such unbalanced configurations 
correspond to large values of the effective action and they will not 
contribute to the observables. When closed rings of exchanges occur, 
namely cyclic permutations of bound-pairs, any transient charge imbalance 
should be eliminated. Below a certain temperature, of the order of 
$100 K$ as shown above, the factor $f$ becomes of the order of unity and 
then such coherent multiple exchanges can occur which lead
to the off-diagonal-long-range order. The arguments which lead to this
last part of the picture are very similar to those discussed by Feynman\cite{Feynman2} for the case of liquid helium-4.

Since the partition-function (\ref{path-integral2}) describes a system
of purely boson degrees of freedom, we can gain 
additional support for this picture by discussing what happens in 
a simpler model.
We consider a system of bosons interacting via a Van der Waals
interaction and we add a long-range $\alpha/r$ Coulomb-like interaction
with a very small value for $\alpha$. First let us consider the case
of $\alpha=0$, namely that of a pure self-bound boson 
liquid on a two-dimensional substrate. Such a system has 
a physical realization,
namely, it describes liquid helium-4 on a substrate such as graphite.
The pure liquid neutral helium-4 on a 2D smooth substrate (without 
substrate corrugations) with periodic 
boundary conditions has been studied\cite{PM} using 
path-integral Monte Carlo  simulation\cite{ceperley}. 
It was found that for values of the 2D density below the 2D 
equilibrium density and at low temperature the system forms a 2D liquid 
droplet with non-zero winding number
(superfluid density) for densities well inside the phase separated region.
Let us discuss what we expect to find if we introduce the 
long-range $\alpha/r$ interaction term with a small 
value of $\alpha$ so that a self-bound droplet with at least two 
atoms exists
as a result of the competition of the long-range and the Van der Waals 
attraction. We expect that the uniformity of the system will be spoiled by 
the presence of this
long-range interaction and droplets of certain size will be formed.
The system optimizes the number of atoms $N_0$ in a droplet,
namely, the binding energy per atom will be very small compared to the 
magnitude of the van der Waals interaction.
This implies that an atom in any charged droplet 
will be only very weakly bound to the droplet
and this will allow a non-zero winding number caused from 
exchanges of atoms between these droplets. 
An estimate of the temperature scale where such exchanges 
occur is
\begin{eqnarray}
K_B T_C \sim {{\hbar^2} \over {2 m d^2}}
\end{eqnarray} 
where $d$ is the average distance between neighboring droplets.
We expect to find a similar result if one introduces a weak 
external potential so that the charged-superfluid is 
confined in a charge-stripe-like 
configuration; namely, the off-diagonal-long-range order sets in
at a temperature of the above mentioned magnitude where $d$ is the distance
between successive hole-rich domains.

\section{Concluding remarks}

We have considered the state in which alternating hole-rich and hole-poor 
(antiferromagnetically ordered) domains form in a strongly correlated 
electron system.

Here we have used existing numerical results on the $t-J$ model to 
show that the presence of antiferromagnetically ordered regions
between such hole-rich domains acts as filter which allows only 
pairs of holes to jump from one hole-rich domain to a nearest 
neighboring one at high rates.
This implies that effectively there is a fluid of bound hole pairs
which can move through the entire system. 

We further discussed a  speculative point of view. 
We considered the limit in which pairing correlations within each
hole-rich domain are weak which implies that 
such a domain  is characterized by a small superconductivity ordering
temperature scale.
In this case we argued that because of the existence of the previously
discussed fluid of bound hole-pairs, 
the characteristic energy scale 
associated with phase coherence between such pairs is determined
by the distance between neighboring hole-rich domains and the pair effective
mass inside the antiferromagnetically ordered domains.

\section{Acknowledgements}
We would like to thank B. I. Halperin, C.S. Hellberg 
and S. A. Kivelson for useful discussions and
comments on the original draft of this paper.

\end{document}